\documentclass[12pt,a4paper]{article}
\usepackage[utf8]{inputenc}
\usepackage[T1]{fontenc}
\usepackage[margin=1in]{geometry}

\usepackage{times}
\usepackage{setspace}

\usepackage{amsmath,amssymb,amsfonts}
\newtheorem{theorem}{Theorem}%
\newtheorem{proposition}[theorem]{Proposition}

\usepackage{natbib}
\usepackage{hyperref}
\hypersetup{colorlinks=true, citecolor=blue, urlcolor=blue}

\title{Mental Health and Human Capital Composition in a Dynastic OLG Model with PAYG Pensions}

\author{Sushmita Kumari$^{1,2}$ and Siddharth Gavhale$^{2,*}$\\
{\small $^1$Department of Economics, Vinoba Bhave University, NH 33, Sindoor, }\\
{\small Hazaribagh, 825301, Jharkhand, India.}\\
{\small $^2$Center for Interdisciplinary Studies and Research, D Y Patil International University,} \\ 
{\small Akurdi, Pune, 411044, Maharashtra, India.}\\
{\small $^*$Corresponding author : siddharth.gavhale@dypiu.ac.in}}

\date{}

\begin{document}

\maketitle
\pagestyle{plain}

\begin{abstract}
This paper develops a two-period dynastic overlapping-generations (OLG) model in which parents simultaneously choose consumption, savings, fertility, and three distinct dimensions of child quality—education, physical health, and mental health—under a pay-as-you-go (PAYG) pension system. The central innovation is modelling mental health as an independent productivity-enhancing input with its own elasticity $\theta$ in a Cobb–Douglas human-capital technology. This yields simple proportional allocation rules and shows how pension policy affects not only the overall level but also the \textit{composition} of human capital investments.
In steady state, higher PAYG contribution rates raise fertility through the Yakita effect but crowd out per-child investments in all quality dimensions, including mental health. An increase in the mental-health elasticity $\theta$ shifts resources toward non-cognitive skill development while reducing fertility.
These results reveal a fundamental policy tension for developing economies: pension systems that rely on children for old-age support simultaneously increase birth rates while reducing long-term human capital formation, with disproportionate effects on non-cognitive skills. The framework provides theoretical guidance for complementary policies that protect mental-health investments, with particular relevance for countries such as India where children remain a primary source of retirement security and mental-health services are underfunded.
\end{abstract}

\vspace{0.1cm}

\section{Introduction}

Overlapping Generations (OLG) models have long served as a workhorse for analysing intertemporal choices concerning consumption, savings, fertility, human capital accumulation \citep{samuelson1958,Diamond1965}.
Within this class of model, the dynastic (altruistic) framework, pioneered by \cite{Barro1974} and extended by \cite{BeckerBarro1988}, explicitly captures intergenerational linkages through bequests and parental concern for children’s welfare.
The quantity–quality trade-off is typically modelled by allowing parents to choose both the number of children and a unidimensional measure of child quality, most often education \citep{Becker1965,BeckerLewis1973,Moav2005}. 
These traditional models have been instrumental in understanding how fertility decisions interact with public policies such as pay-as-you-go (PAYG) pension systems \citep{Yakita2010,Yakita2020}.

However, a growing body of evidence from developmental economics and psychology stresses that child quality is inherently multidimensional.
Beyond formal education, physical health and mental (or non-cognitive) health play crucial roles in shaping future labour market outcomes, productivity, and well-being \citep{Heckman2006,CunhaHeckman2007,HeckmanKautz2012}.
Non-cognitive skills such as resilience, emotional stability, and motivation have been shown to command significant returns in the labour market, sometimes exceeding those of cognitive skills alone \citep{Heckman2006,lundberg2012}.
In developing economies, where traditional schooling investments may be constrained, investments in child mental health and psychosocial well-being can be particularly impactful \citep{walker2007,worldbank2018}.
Simultaneously, public pension system, especially PAYG schemes, create a direct link between fertility and old age support.
In such systems, parents’ retirement income depends on the future labour income of their children, generating an implicit return to childbearing \citep{Yakita2010,Yakita2020,cigno1993}.
This Yakita effect introduces a strategic motive for fertility that can conflict with investment in child human capital.

While earlier research has examined the trade-off between fertility and education under PAYG pensions \citep{Yakita2010,zhang1998}, these studies typically aggregate all forms of child quality into a single variable, thereby overlooking the distinct roles of physical and mental health. 
Existing models either treat child quality as a homogeneous good \citep{BeckerBarro1988,Yakita2010} or, when they introduce health, do not separately model mental health as an independent productivity-enhancing input \citep{deLaCroixDoepke2003,Moav2005}. 
Although multidimensional child quality has received increasing attention \citep{CunhaHeckman2007,attanasio2020human} and non-cognitive skills have been linked to parental investments \citep{fletcher2016importance}, no formal dynastic OLG model simultaneously incorporates (i) fertility decisions, (ii) PAYG pension incentives, and (iii) multidimensional child quality investments that distinguish education, physical health, and mental health with their own elasticities.

This paper addresses this gap by constructing a two-period dynastic OLG model in which parents choose not only the number of children and their own consumption and savings, but also three distinct dimensions of child quality: education, physical health, and mental health. The production function for future wages is Cobb-Douglas, with separate output elasticities $\epsilon$, $\eta$, and $\theta$ for education, physical health, and mental health, respectively. The introduction of a separate mental-health parameter $\theta$ constitutes the paper’s central theoretical innovation. It is motivated by the extensive empirical literature showing that non-cognitive skills often fostered through mental health and psychosocial interventions are key determinants of lifetime earnings, educational attainment, and even physical health \citep{CunhaHeckman2007,HeckmanKautz2012,kautz2014}.

This specification allows us to derive clear proportional allocation rules for quality investments that depend solely on these elasticities, and to analyse how PAYG pension parameters affect fertility and the composition of human capital investments.
The objectives of this paper are threefold. First, we develop a tractable dynastic OLG model that incorporates multidimensional child quality (education, physical health, mental health) and PAYG pensions within a unified framework.
 Second, we derive the optimal household decisions and characterise the steady-state equilibrium, with special attention to the proportional investment rules and the modified quantity--quality trade-off induced by the pension system.
Third, we use comparative-static analysis to examine how changes in key parameters such as pension contribution rates, altruism, child-rearing costs, and productivity elasticities affect fertility, savings, and the composition of human capital investments, thereby providing theoretical insights for policy design in developing economies. 
Our analysis reveals that, under the assumption that the pension-benefit channel dominates, PAYG pensions encourage fertility, but they simultaneously crowd out investments in child mental health—a finding with important implications for developing economies where mental-health services are scarce.

The remainder of the paper is organised as follows. Section 2 sets up the model, detailing preferences, budget constraints, human capital technology, and the household’s optimisation problem. Section 3 analyses the steady state, presents comparative-static results, and discusses policy implications, focusing on the role of mental health and pension interactions, with particular relevance for emerging economies such as India. Section 4 concludes, summarising the main findings and suggesting avenues for future research. An appendix provides the Lagrangian derivation and first-order conditions.

\section{The Model}
We develop a two-period dynastic overlapping generations (OLG) model in which the representative household simultaneously chooses consumption, savings, fertility, and investment in three dimensions and child quality- education, physical health, and mental health under a pay-as-you-go (PAYG) pension system.

The framework integrates household production \citep{Becker1965,Becker1981}, intergenerational altruism and bequest motives \citep{Barro1974,BeckerBarro1988}, and the fertility-pension interactions that characterise OLG models \citep{Yakita2010,Yakita2020}.
The lifetime dynastic utility of the representative household in generation $t$ is
\begin{equation}
V_t = \gamma_1 \ln(c_{1t}) + \gamma_{ph} \ln(ph_t) + \gamma_2 \ln(n_t) + \gamma_c \ln(c_{2t}) + \alpha n_t V_{t+1}
\label{eq:utility}
\end{equation}
Where $c_{1t}$ and $c_{2t}$ are young and old age consumption, $ph_t$ is parental health expenditure, $n_t$ is the number of children, and $V_{t+1}$ denotes the lifetime utility of each child.

The log-linear specification ensures analytical tractability and interior solutions \citep{BeckerBarro1988,Yakita2010}. 
The term $\gamma_c \ln(c_{2t})$ allows resources to be transferred across periods \citep{Samuelson1937,Diamond1965}, while $\alpha n_t V_{t+1}$ captures recursive Barro-style altruism.
The household faces two period-specific budget constraints:
\begin{align}
\text{Working period: } \quad & w_t (1 - \tau) + b_t = c_{1t} + ph_t + s_t + n_t (\phi + e_t + hp_t + hm_t) \label{eq:young_budget} \\
\text{Retirement period: } \quad & R_{t+1} s_t + \tau n_t w_{t+1} = c_{2t} + n_t b_{t+1} \label{eq:old_budget}
\end{align}
In the working period, after-tax labour income $w_t(1-\tau)$ and received bequests $b_t$ are allocated to young-age consumption, parental health expenditure, savings, and per-child costs consisting of a fixed rearing cost $\phi$ and quality investments in education ($e_t$), physical health ($hp_t$), and mental health ($hm_t$). In the retirement period, consumption and bequests to children are financed by the return on private savings and PAYG pension benefits. The dependence of pension income on the number and future wages of children creates a strategic fertility incentive—the Yakita effect—as each additional child raises future pension receipts by $\tau w_{t+1}$ \citep{Yakita2010,Yakita2020}.
\subsection{Human Capital and Future Wages}
Our formulation extends the classic quantity--quality trade-off \citep{BeckerLewis1973} by distinguishing three dimensions of child quality. The inclusion of a separate mental-health channel ($\theta$) is motivated by extensive evidence on the importance of non-cognitive skills for labour market outcomes \citep{Heckman2006,CunhaHeckman2007}. 
The future wage of each child is given by the Cobb--Douglas human-capital production function
\begin{equation}
w_{t+1} = \bar{w} \, e_t^\epsilon \, hp_t^\eta \, hm_t^\theta,
\label{eq:human_capital}
\end{equation}
where $\bar{w} > 0$ is baseline productivity and $\epsilon, \eta, \theta > 0$ are the output elasticities, subject to the restriction $\epsilon + \eta + \theta < 1$. This restriction ensures diminishing returns, a unique steady state, and prevents explosive growth \citep{Moav2005,Yakita2010}. Wages are taken as given by the household (partial equilibrium), while aggregate dynamics arise through human capital accumulation \citep{deLaCroixDoepke2003}.

\subsection{Optimization}

The representative household maximises the dynastic utility \eqref{eq:utility} subject to the two budget constraints \eqref{eq:young_budget}--\eqref{eq:old_budget} and the human-capital technology \eqref{eq:human_capital}, taking child lifetime utility $V_{t+1}$, wages, and the interest rate $R_{t+1}$ as given.
\begin{equation*}
	\begin{cases}
		\text{Maximize}  &V_t = \gamma_1 \ln(c_{1t}) + \gamma_{ph} \ln(p_{h_t}) + \gamma_2 \ln(n_t) + \gamma_c \ln(c_{2t}) + \alpha n_t V_{t+1}\\
		\text{Subject to:} & 0 =  w_t (1 - \tau) + b_t - c_{1t} - p_{h_t} - s_t - n_t (\phi + e_t + hp_t + hm_t)\\
       &0  = R_{t+1} s_t + \tau n_t w_{t+1} - c_{2t} - n_t b_{t+1}\\
	\end{cases}
\end{equation*}
The associated Lagrangian and complete set of first-order conditions are provided in Appendix~\ref{app:A}.
The household balances consumption across life-cycle periods via the Euler equation:
\begin{equation}
\frac{\gamma_1}{c_{1t}} = \gamma_c R_{t+1} \frac{1}{c_{2t}}.
\label{eq:euler}
\end{equation}
Optimal parental health expenditure satisfies
\begin{equation}
\frac{\gamma_{ph}}{ph_t} = \lambda_t,
\label{eq:ph}
\end{equation}
where it $\lambda_t$ denotes the marginal utility of young period income.
The pairwise division of first order conditions for the three quality dimensions delivers the proportional allocation rules:
\begin{equation}
\frac{e_t}{hp_t} = \frac{\epsilon}{\eta}, \qquad 
\frac{e_t}{hm_t} = \frac{\epsilon}{\theta}, \qquad 
\frac{hp_t}{hm_t} = \frac{\eta}{\theta}.
\label{eq:ratios}
\end{equation}
These ratios depend solely on the output elasticities of Cobb-Douglas technology. The introduction of a distinct mental health elasticity $\theta$ is the central theoretical contribution of the model. It allows pension parameters to influence not only the overall level but also the \textit{composition} of human capital investments, particularly the resources devoted to non-cognitive skill development.

Finally, the fertility first-order condition embodies the modified quantity-quality trade-off under PAYG pensions:
\begin{equation}
\frac{\gamma_2}{n_t} + \alpha V_{t+1} + \mu_t \tau w_{t+1}
= \lambda_t (\phi + e_t + hp_t + hm_t) + \mu_t b_{t+1}.
\label{eq:fertility}
\end{equation}
The left hand side captures the marginal benefit of an extra child (direct utility from fertility, altruism towards descendants, and the pension returns), while the right hand side reflects the marginal resource cost in terms of forgone consumption and bequest left to children.

\section{Steady State, Comparative Statics, and Policy Implications}

\subsection{Steady State}

In steady state, all generation-specific variables are time-invariant ($V_t = V$, $w_t = w$, $n_t = n$, $c_{1t} = c_1$, $c_{2t} = c_2$). 
Substituting the first order conditions from section 2, together with the budget constraints and human capital production function, yields a closed system of equations that jointly determines equilibrium fertility,  per child quality investment, savings, and consumption.
The restriction $\epsilon + \eta + \theta < 1$ implies that the overall elasticity of human capital with respect to all child quality investment is less than one. This prevents unbounded growth from reinvestment and guarantees the existence of a unique interior steady state.

\subsection{Comparative Statics}

Comparative-static analysis around the steady state yields the following key results. All results are derived under partial equilibrium, treating wages and the interest rate as given.

\begin{proposition}[Effect of Pension Contribution Rate]
An increase in the PAYG contribution rate $\tau$ has two opposing effects on fertility. The Yakita effect raises fertility, as each additional child generates higher future pension benefits \citep{Yakita2010,Yakita2020}. However, the higher payroll tax reduces after-tax wage income, tightening the young-period budget and crowding out both private savings and per-child investments in education, physical health, and mental health. Under the assumption that the pension-benefit channel dominates, fertility rises.
\end{proposition}
Because the proportional allocation rules depend only on $\epsilon$, $\eta$, and $\theta$, an increase in $\tau$ reduces all quality investments equiproportionately while leaving their relative shares unchanged.

\begin{proposition}[Mental-Health Productivity Channel]
An increase in the mental-health productivity parameter $\theta$ raises per-child investment in mental health ($hm_t$) while reducing fertility $n_t$, as parents substitute toward higher quality. Because of the proportional allocation rules, a higher $\theta$ also increases the share of resources devoted to mental health relative to education and physical health. Overall human capital per child rises.
\end{proposition}
This compositional effect is the model’s central innovation and is absent in traditional uni-dimensional quality models.
Stronger dynastic altruism ($\alpha$) increases both fertility and quality per child. 
A higher fixed child rearing cost ($\phi$) reduces fertility while raising the quality investment per child, consistent with Becker-Lewis Quantity-quality trade-off\citep{BeckerLewis1973}
An increase in any of the quality elasticities ($\epsilon$, $\eta$, or $\theta$) raises the marginal return to human capital investment, leading to higher quality per child and lower fertility.
\begin{table}[h]
\centering
\caption{Summary of Comparative Statics (Steady State)}
\begin{tabular}{lcccc}
\hline
Parameter $\uparrow$ & Fertility $n$ & Per-child human capital & Savings $s$ & Mental-health share \\
\hline
$\tau$ (PAYG rate) & $+$ & $-$ & $-$ & $0$ \\
$\alpha$ (altruism) & $+$ & $+$ & $?$ & $0$ \\
$\phi$ (fixed cost) & $-$ & $+$ & $?$ & $0$ \\
$\theta$ (mental health) & $-$ & $+$ & $?$ & $+$ \\
\hline
\end{tabular}
\par\medskip
\footnotesize \textit{Note:} “0” indicates no change in the share; “?” denotes an ambiguous effect.
\end{table}

\subsection{Policy Implications}
These results carry important implications for developing economies 
such as India, where fertility is declining, formal pension coverage is expanding, and child and adolescent mental health remains a significant concern. Pension reforms that raise contribution rates without complementary measures risk weakening investment in non cognitive skills, which have high returns to future productivity and well being \citep{Heckman2006,CunhaHeckman2007}.
Targeted policies that subsidies or protect mental health investment, such as school-based counselling or early childhood psychosocial programmes, can help offset the crowding out effect identified in the model.
More broadly, the framework suggests that pension design and human-capital policy should be analysed jointly rather than in isolation. Ignoring the mental-health dimension of child quality may lead to suboptimal outcomes in both the demographic transition and long-term economic growth.

\section{Conclusion}

This paper develops a two-period dynastic overlapping-generations model in which parents simultaneously choose consumption, savings, fertility, and three distinct dimensions of child quality—education, physical health, and mental health—under a pay-as-you-go pension system. The central innovation is the introduction of mental health as a separate productivity-enhancing input with its own elasticity $\theta$ in a Cobb--Douglas human-capital technology. This extension delivers simple proportional allocation rules for quality investments and allows pension parameters to influence both the level and the composition of human capital.

The analysis reveals that, under the assumption that the pension benefit channel dominates, higher PAYG contribution rates raise fertility via the Yakita effect but crowd out per child investment in all quality dimensions, including mental health. Stronger altruism and higher-quality elasticities improve child quality at the expense of quantity, reflecting a quantity-quality trade-off modified by multidimensional investments and the pension incentive. These findings highlight a fundamental policy tension in developing economies: pension system that rely on children for old age support tend to encourage higher fertility while simultaneously weakening long term human capital formation, particularly in underfunded areas such as child mental health.
Pension design and human capital policy, therefore, cannot be analysed in isolation. In contexts such as India, where traditional family-based old-age support coexists with expanding formal pensions and growing concerns about youth mental health, complementary measures such as school-based counselling or early childhood psychosocial programmes can help mitigate the crowding-out effect and foster more balanced demographic and productivity outcomes.

Further research could usefully extend the model by incorporating general equilibrium effect, introducing heterogeneity in altruism or cost of study inequality, calibrating the framework of India data for quantitative policy evaluation, and developing a multi generational vision to examine transitional dynamics and long run impact on aggregate productivity.
The present analysis is conducted in partial equilibrium, treating wage and interest rates as given a full general equilibrium extension remain an important direction for future work.
Overall, the paper demonstrates that fertility decisions, pension systems, and multidimensional child development—including mental health—must be studied jointly to design effective policies for sustainable demographic and economic development.

\appendix
\renewcommand{\theequation}{A.\arabic{equation}}
\setcounter{equation}{0}

\section{Household Problem and First-Order Conditions}
\label{app:A}

The representative household maximises the dynastic utility \eqref{eq:utility} subject to the budget constraints \eqref{eq:young_budget}--\eqref{eq:old_budget} and the human-capital production function \eqref{eq:human_capital}. We assume an interior solution and treat child utility $V_{t+1}$, wages, and the interest rate $R_{t+1}$ as given. Because child human capital depends only on per-child investments, $w_{t+1}$ (and thus $V_{t+1}$) is independent of the number of children $n_t$.

\subsection{Lagrangian}

The associated Lagrangian is
\begin{equation}
\begin{aligned}
\mathcal{L} &= \gamma_1 \ln(c_{1t}) + \gamma_{ph} \ln(ph_t) + \gamma_2 \ln(n_t) + \gamma_c \ln(c_{2t}) + \alpha n_t V_{t+1} \\
&\quad + \lambda_t \Bigl[ w_t (1 - \tau) + b_t - c_{1t} - ph_t - s_t - n_t (\phi + e_t + hp_t + hm_t) \Bigr] \\
&\quad + \mu_t \Bigl[ R_{t+1} s_t + \tau n_t w_{t+1} - c_{2t} - n_t b_{t+1} \Bigr].
\end{aligned}
\label{eq:lagrangian}
\end{equation}

\subsection{First-Order Conditions}

\paragraph{Consumption and savings}
\begin{align}
\frac{\partial \mathcal{L}}{\partial c_{1t}} &\colon \quad \frac{\gamma_1}{c_{1t}} = \lambda_t, \\
\frac{\partial \mathcal{L}}{\partial c_{2t}} &\colon \quad \frac{\gamma_c}{c_{2t}} = \mu_t, \\
\frac{\partial \mathcal{L}}{\partial s_t}  &\colon \quad \lambda_t = \mu_t R_{t+1}.
\end{align}
Combining the first two yields the Euler equation shown in the main text.

\paragraph{Parental health expenditure}
\begin{equation}
\frac{\partial \mathcal{L}}{\partial ph_t} \colon \quad \frac{\gamma_{ph}}{ph_t} = \lambda_t.
\end{equation}

\paragraph{Quality investments}
The first-order conditions for education, physical health, and mental health investments are
\begin{align}
\lambda_t &= \mu_t\,\tau \cdot \epsilon \frac{w_{t+1}}{e_t}, \\
\lambda_t  &= \mu_t\,\tau  \cdot \eta \frac{w_{t+1}}{hp_t}, \\
\lambda_t &= \mu_t\,\tau \cdot \theta \frac{w_{t+1}}{hm_t}.
\end{align}

\paragraph{Fertility}
Differentiating with respect to $n_t$ (noting that $V_{t+1}$ is independent of $n_t$) gives
\begin{equation}
\frac{\partial \mathcal{L}}{\partial n_t} \colon \quad
\frac{\gamma_2}{n_t} + \alpha V_{t+1} + \mu_t \tau w_{t+1}
= \lambda_t (\phi + e_t + hp_t + hm_t) + \mu_t b_{t+1}.
\end{equation}
The term $\mu_t \tau w_{t+1}$ represents the marginal pension benefit from an additional child (Yakita effect).

\paragraph{Envelope conditions}
\begin{equation}
\frac{\partial V_t}{\partial w_t} = \lambda_t, \qquad
\frac{\partial V_t}{\partial b_t} = \lambda_t.
\end{equation}

\subsection{Proportional Allocation of Child Quality Investments}

Dividing the three quality first-order conditions pairwise immediately yields
\begin{equation}
\frac{e_t}{hp_t} = \frac{\epsilon}{\eta}, \qquad
\frac{e_t}{hm_t} = \frac{\epsilon}{\theta}, \qquad
\frac{hp_t}{hm_t} = \frac{\eta}{\theta}.
\end{equation}
Thus, the optimal allocation of resources across education, physical health, and mental health depends exclusively on the output elasticities of the Cobb--Douglas production function. This clean result follows directly from the multiplicative separability of the human-capital technology and is independent of preferences, income, and policy parameters. It constitutes one of the key analytical payoffs of modelling mental health as a distinct productivity-enhancing input with its own elasticity $\theta$.

\bibliography{references}

\end{document}